# Outreach Strategies for Vaccine Distribution: A Multi-Period Stochastic Modeling Approach

Yuwen Yang, Jayant Rajgopal[1]

Department of Industrial Engineering, University of Pittsburgh, Pittsburgh, PA 15261, USA

**Acknowledgement:** This work was partially supported by the National Science Foundation via Award No. CMII-1536430.

---

[1] Corresponding author; e-mail: rajgopal@pitt.edu; Tel. +1 412 624 9840; ORCID 0000-0001-7730-8749

# Outreach Strategies for Vaccine Distribution: A Multi-Period Stochastic Modeling Approach

**Abstract**

Vaccination has been proven to be the most effective method to prevent infectious diseases. However, in many low and middle-income countries with geographically dispersed and nomadic populations, last-mile vaccine delivery can be extremely complex. Because newborns in remote population centers often do not have direct access to clinics and hospitals, they face significant risk from diseases and infections. An approach known as outreach is typically utilized to raise immunization rates in these situations. A set of these remote locations is chosen, and over an appropriate planning period, teams of clinicians and support personnel are sent from a depot to set up mobile clinics at these locations to vaccinate people there and in the immediate surrounding area. In this paper, we model the problem of optimally designing outreach efforts as a mixed integer program that is a combination of a set covering problem and a vehicle routing problem. In addition, because elements relevant to outreach (such as populations and road conditions) are often unstable and unpredictable, we address uncertainty and determine the worst-case solutions. This is done using a multi-period stochastic modeling approach that considers updated model parameter estimates and revised plans for subsequent planning periods. We also conduct numerical experiments to provide insights on how demographic characteristics affect outreach planning and where outreach planners should focus their attention when gathering data.

# 1. Introduction

As a biological preparation against infectious disease, vaccines have averted 2 to 3 million deaths annually [1], and coverage rates have improved significantly over the years under the guidance of the World Health Organization's Expanded Programme on Immunization (WHO-EPI) and the Global Alliance for Vaccines and Immunization (Gavi) [2,3]. However, in many of the poorest countries, getting childhood vaccines delivered to their final destinations can be an extremely complex process. Although many low and middle-income countries (LMICs) can often obtain vaccines at low cost, operating a vaccine distribution system can be a challenge. Many vaccines require a narrow temperature range of between 2 and 8°C during storage and transportation, which in turn brings with it high distribution and storage costs. In addition to the challenge of planning for storage devices and transportation capabilities to distribute vaccines throughout the country, geographically dispersed or nomadic populations also present a major challenge. As a result, in many countries significant portions of the population have no direct access to health clinics.

Inadequate infrastructure and geographic barriers such as poor road conditions or limited access to transportation can further compound this problem. For example, in Niger, around 90% of the roads are not paved [4]. A recent study published in The Lancet Global Health estimated that across 48 sub-Saharan countries, 28.2% of women of child-bearing age are more than 2 hours travel time (combined walking and motorized) away from the nearest hospital [5]. The study also found wide variations with the percentage ranging from under 25% in South Sudan to over 90% in several countries including Nigeria, Kenya, Swaziland and Burundi. Another recent study in Uganda [6] concluded that difficulty in access to immunization centers due to poor road terrain has a significant effect and results in low immunization coverage. Thus, people living in remote locations in LMICs often face significant difficulty in obtaining routine vaccinations, and the WHO estimates that almost 20 million infants worldwide are at high risk from vaccine-preventable diseases such as polio, measles, yellow fever and tuberculosis [7].

To supplement the fixed vaccine distribution network, an approach known as *outreach* is typically utilized to raise immunization rates, especially in remote areas where direct access to clinic services is limited or unavailable. Clinicians and support personnel are sent from an existing (permanent) clinic location to render vaccination services at one or more of these remote population centers (villages, communities, settlements, etc.). While the exact terminology varies from country to country, we will refer to the permanent facility as a *depot* and the temporary facility at a remote location as a *mobile clinic*. People at the population center where the mobile clinic is located and other population centers that are within a reasonable distance from it attend the clinic to get vaccinated. Note that this service is distinct from a *campaign* (a one-time attempt), in that outreach is periodic and repeated at regular time intervals over successive planning periods; these might range from 1 month to 6 months in different countries.

Mobile clinics can offer more flexibility and viability when treating vulnerable and isolated populations [8,9] and avoid unnecessary fixed facility, inventory, and labor cost [10]. Furthermore, outreach is proven to dramatically raise the overall immunization rates in resource-deprived countries that suffer from extremely low coverage rates. An early study in Kenya estimated that outreach increased the coverage rate in the lowest density zone in Kenya from 25% to 57% and from 54% to 82% in the area with greatest population density [11]. With the support of the WHO, outreach activities encompassing 1,982 mobile clinics and 5,964 personnel were able to cover 80% of targeted infants in September 2015 in Yemen; 290,498 children were vaccinated by these actions; and in 2018, 44 mobile clinics were set up to serve populations to hard-to-reach areas in Syria [8].

While there has been some relatively recent work on the network design phase of the WHO vaccine distribution chain [12–16], outreach has received virtually no attention in the academic literature and there are almost no quantitative models available to help decision makers create a robust outreach strategy. In this paper we propose a general model for LMICs to develop outreach plans at a given fixed clinic location. We also use a multi-period stochastic modeling approach to address uncertainty and update operational plans in each planning period based on updated information. Numerical experiments are conducted to develop insights for outreach planners on relationships between demographic characteristics and uncertainty in input parameter estimates.

Section 2 provides some background along with definitions and our assumptions, and also reviews prior work that is directly relevant to our problem. We then provide an MIP model formulation in Section 3 for the initial planning period. An extension of the model with uncertainty considerations is presented in Section 4 for use in subsequent periods. We introduce the notion of "value of information" in Section 5 and present numerical results in Section 6 to gain insight on the effect of different aspects of uncertainty. Finally, we discuss the results and summarize our findings in Section 7.

## 2. Problem Description and Review of Relevant Prior Work

While outreach has been proven to be effective at increasing vaccination rates in resource-deprived regions of the world, there is no standard structure or process that every country follows. A typical process might be one where a medical team departs from an existing district medical center or clinic (i.e., the depot) in a van or truck, carrying supplies and vaccines in cold boxes. The team then sets up a mobile clinic at a remote location and vaccinates the residents of that location as well as residents from nearby locations. If multiple locations are visited on an outreach trip, the team would drive to the locations sequentially to provide service at each, before returning to the original depot at the end of the day. However, each country has its own outreach policy and approach to conduct outreach. For example, an outreach team might consist of clinicians who come from one location (usually, the depot) and other staff members who might come from a different location. In some cases, the vaccines might be delivered by a separate logistical team and stored in a refrigerator at some suitable facility at the remote location a day or two before the clinical team arrives there to set up the mobile clinic. In some countries, it might be possible for the team to stay overnight at a remote location, so that the team could conduct more mobile clinics in the course of an outreach trip. In general, unlike with the operation of fixed clinics and the associated vaccine distribution system, there are no clear standards on how outreach should be done.

Despite significant variations in economy, geography, demography, etc., and thus, in how outreach is done across all these countries, this paper aims to provide a relatively rigorous process for outreach trips across all countries to meet the WHO's goal of providing the entire targeted population with the opportunity to be vaccinated. We present a mixed integer programming (MIP) formulation to optimize outreach strategies, and the structure we assume and describe next represents a very common one we have found, based on conversations with country-level public health professionals, and some of the non-governmental organizations involved with outreach.

We begin by defining a *location* as a population center (typically, an aggregation of people, such as a village, settlement or community) that is targeted for outreach from the depot. A location is *covered* either by a mobile clinic at the location itself, or by a mobile clinic at a nearby location to which it could be *assigned*. We define a *planning period* as a given interval of time (typically, 3 months or 6 months) over which every location must be covered once by outreach. To retain a tractable MIP model while accounting for the various associated complexities and diversity as best we can, we include three sets of decisions into our consideration. The first is choosing the locations for mobile clinics as a subset of all defined locations. The second set of considerations addresses how to assign the remaining locations to mobile clinics. Note that a location can be assigned to a mobile clinic only if it is within the *maximum coverage distance* (MCD) to the location where the mobile clinic is to be conducted. The MCD can be defined as the maximum distance that people must travel to get vaccinated and is determined by the planners in the country. A mobile clinic could thus cover multiple locations. We ensure that each location is assigned to a specific mobile clinic or to the depot if it is within the MCD from it. Third, we determine an optimal set of outreach trips that ensure that all locations where mobile clinics are to be conducted are visited once within the planning period. Each outreach trip would use a vehicle to carry the required clinical and support personnel and equipment along with the required amount of vaccine for the location(s) to be covered by the trip. Once vaccination is done and the team leaves the location, the mobile clinic is abolished until the time the outreach team visits that location during the next planning period.

Within a planning period, multiple outreach trips can be undertaken but there can be only one ongoing outreach trip at any given time because the depot typically has access to only one vehicle. Each outreach trip must depart from the depot and return to the depot after it visits one or more locations where mobile clinic are conducted. The vehicle utilized in an outreach trip is typically a truck or a van with several coolers or cold boxes and is thus capacitated in terms of how much vaccine can be carried. Note that the vaccine regimens are not identical across countries, and based on the total demand that is expected at locations to be covered and the vaccine vial volumes, we can estimate the total volume associated with expected demand at each location during each planning period.

To model a realistic process for outreach, we define a *maximum trip duration* (MTD) for each trip. This might be 8 to 12 hours if all personnel need to return to the depot on the same day; in the case that they could stay overnight at a location where a mobile clinic is conducted, the MTD could be longer. We also define the *service time* as sum of the time used to set up and dismantle the mobile clinic and the time allocated to vaccinate targeted population members who come to the clinic. Different clinics need not have identical service times; a clinic at a location that is assigned to cover a larger population is likely to have a longer service time. The service time at the originating depot can be set, as appropriate, either to zero or to the actual time required to load vaccines and prepare the team on the day of the trip, and any additional time at the end of the trip. The travel times between the depot and the stop(s) during an outreach trip are obtained by dividing the corresponding distances by the average vehicle speed (e.g., 25 km/h).

We consider two components of cost in our objective function. The first is the direct cost associated with running a mobile clinic at a remote location. This cost includes the setup cost for clinic, the cost of renting or obtaining space, any labor costs specific to the onsite location, potential storage and energy consumptions costs there, and any other local cost. The second cost component is the trip-related cost that is assumed to be proportional to the duration of the trip. This might include fuel costs, vehicle depreciation, hourly wages/allowances paid to the team and driver for the trip and travel, vehicle rental costs, etc. We assume this results in an average cost per hour that is used to compute the cost of the trip based on its planned duration. The total cost is thus determined by the locations of the mobile clinics and the routes taken by the depot's vehicle on its outreach trips. Since the fixed costs for the depot are independent of our outreach planning decisions, they are not part of the optimization problem, and we also ignore any possible costs associated with patients traveling from their location to the mobile clinic. Our objective is to minimize the overall cost for outreach in a planning period, while guaranteeing universal access; this is in alignment with the WHO's goal of ensuring that the entire population is covered, and the goal of local governments that this be done as economically as possible.

In summary, we have the following assumptions:

1) Our objective is to minimize the sum of direct mobile clinic costs and outreach trip costs.
2) Locations for mobile clinics as part of outreach are chosen from a set of existing locations (targeted population centers).
3) A location is said to be *covered* by a mobile clinic (or the depot) if it is within the specified MCD.
4) A location can be *assigned* to a mobile clinic only if it is covered by that mobile clinic. Each location is assigned to one mobile clinic, so that the entire population to be covered by the depot has the opportunity to be vaccinated.
5) Each depot is assumed to have a single vehicle that is available for outreach. In each outreach trip, the vehicle departs from the depot, visits one or more locations to conduct mobile clinics, and returns to the depot.
6) Multiple outreach trips are made within a planning period, and every location where a mobile clinic is to be conducted must be visited once within the planning period by an outreach trip.

7) There is a service time associated with each mobile clinic and a travel time between locations where mobile clinics are held. We assume that the entire trip, including travel times and service time(s), must be completed within the MTD.

8) The vehicle is capacitated, and we assume the capacity is more than what is required at any single location.

The MCD in assumption 3 is assumed to be set to a value that is acceptable in the country being considered. Assumption 4 captures the WHO policy of ensuring that every child has the opportunity to be vaccinated. Assumptions 5, 6 and 7 derive from the structure we assume based on common practice, and assumption 8 is required to ensure feasibility.

Given the above description of the problem and its structure, we now review the relevant literature. To our knowledge there are only two prior attempts at even approaching outreach from a quantitative standpoint. Lim et al. [17] were the first to address outreach analytically, and they solved a static design problem to select the best set of locations for mobile clinics to be served by a single depot under a fixed budget, with the objective of maximizing the number of individuals vacccinated via outreach. Although their approach considers budget limitations, which is not uncommon in LMICs, a shortcoming of this work is that it does not align well with the WHO's goal of universal access. Moreover, the main focus of the paper was to study how many people could be vaccinated under different assumptions on coverage based on the distance that patients might need to travel to get to the mobile clinic. The authors contrast various models of coverage for their problem using data derived from the state of Bihar in India.

In more recent work, Mofrad [18] also considers a static design problem, but one that is more comprehensive, in that it studies the tradeoff between fixed and mobile clinics to find how many of each to locate in a large area (such as a district or a region), and where to locate these. She proposes a mixed integer programming model to minimize costs under uncertain demand, and presents illustrative results for the case where demand follows a log-normal distribution. This work is theoretically rigorous and also attempts to provide some insights with respect to how demand uncertainty affects location decisions. However, the model presented is a stylized one, and based on the results from a limited set of artificial test problems that are solved, it is hard to draw too many general conclusions.

A shortcoming of both of the publications discussed above is that they look at a static design problem; neither one studies the problem on an ongoing basis. While locations of depots are one-time decisions, outreach from a depot is typically done regularly, but at potentially different locations and different points in time during each planning period. Because the same plan is not followed each time, the underlying model parameters and solution need to be updated repeatedly. In this paper we assume that the locations of the fixed clinics are given, and we consider the design and operation of outreach programs from a specific fixed clinic (depot). A mixed integer program provides the overall design of the outreach effort along with the operating plan for the initial planning period, and we use a multi-period stochastic modeling approach that considers updated bounds on uncertain parameters and obtains revised plans for subsequent planning periods. We also conduct a numerical study to provide management insights on the relationship between demographic characteristics and uncertain parameters, and guidance on where to focus attention while gathering data to estimate these parameters.

The proposed problem under these assumptions can be viewed as a combination of a set covering problem (SCP) and a vehicle routing problem (VRP): the process of choosing locations for mobile clinic and assigning other locations to each clinic can be viewed as an SCP, while the routes to visit these locations can be viewed as a VRP. A typical SCP in this context would choose the optimal facility locations with the objective of minimizing cost or maximizing the total demand covered [19,20]. This is a well-studied problem in the operations research community and has been widely applied in the heath care area [10]. The VRP, due to its wide application and importance in distribution networks, has also been widely studied by researchers; the goal is to obtain an optimal vehicle trip strategy to serve a set of customers. However, due to its complexity, exact algorithms such as branch-and-cut and branch-and-price usually have a size limit of 50 to 100 nodes; the problem is thus often solved by approximation algorithms and heuristics to find high quality solutions [21–24]. Extensions of the VRP include ones where there are predefined time-related constraints during a vehicle trip, such as specific time windows within which customers must be served [25], time windows at the depot within which a vehicle must depart from or arrive at the depot [26], or constraints on the maximum trip duration [26–28]; our formulation considers the latter.

## 3. Model Formulation

*Parameters*:

$n$: Total number of targeted locations

$i$: Index of locations, $1 \leq i \leq n$ and $i = 0, n+1$ if $i$ is the depot

$k$: Index of outreach trips

$b_i$: Volume of vaccine demanded at location $i$ over the planning period

$f_i$: Fixed cost of running a mobile clinic at location $i$

$c$: Average transportation cost per hour

$d_{ij}$: Distance between location $i$ and location $j$ (with $d_{ii} = 0$)

$D$: Maximal coverage distance (MCD)

$a_{ij} \in \{0,1\}$: 1 if location $i$ is within a distance $D$ from location $j$, 0 otherwise

$K$: Maximum number of outreach trips that can be made within the planning period

$t_{ij}$: Travel time from location $i$ to location $j$

$s_i$: Service time at location $i$

$r$: Maximum trip duration (MTD)

$p$: Capacity of vehicle

*Variables*:

$X_{ij} \in \{0,1\}$: 1 if location $j$ is assigned to mobile clinic at location $i$, 0 otherwise

$Y_i \in \{0,1\}$: 1 if there is a mobile clinic at location $i$, 0 otherwise

$Z_{ijk} \in \{0,1\}$: 1 if location $i$ is followed by location $j$ in outreach trip $k$, $k \leq K$

$U_{ik}$: Cumulative vaccine volume distributed by outreach trip $k$ when leaving location $i$ after conducting a mobile clinic there, $k \leq K$

$W_i$: Total volume of vaccine sent to mobile clinic at location $i$

MIP-1:

$$Min \sum_{1 \leq i \leq n} f_i Y_i + \sum_{0 \leq i \leq n+1} \sum_{0 \leq j \leq n+1} \sum_{1 \leq k \leq K} ct_{ij} Z_{ijk} \qquad (1)$$

*subject to*

$$X_{ij} \leq a_{ij} \qquad \forall i,j \quad (2)$$

$$X_{ij} \leq Y_i \qquad 0 \leq i \leq n, 1 \leq j \leq n \quad (3)$$

$$\sum_{0 \leq i \leq n} X_{ij} = 1 \qquad 1 \leq j \leq n \quad (4)$$

$$W_i = \sum_{0 \leq j \leq n+1} b_j X_{ij} \qquad i \leq n \quad (5)$$

$$\sum_{0 \leq j \leq n+1} Z_{0jk} = 1 \qquad \forall k \quad (6)$$

$$\sum_{0 \leq j \leq n+1} Z_{j0k} = 0 \qquad \forall k \quad (7)$$

$$\sum_{0 \leq i \leq n+1} Z_{i(n+1)k} = 1 \qquad \forall k \quad (8)$$

$$\sum_{0 \leq i \leq n+1} Z_{(n+1)ik} = 0 \qquad \forall k \quad (9)$$

$$\sum_{0 \leq j \leq n+1} Z_{ijk} = \sum_{0 \leq j \leq n+1} Z_{jik} \qquad \forall k, 1 \leq i \leq n \quad (10)$$

$$\sum_{0 \leq j \leq n+1} \sum_{1 \leq k \leq K} Z_{ijk} = Y_i \qquad 1 \leq i \leq n \quad (11)$$

$$U_{ik} - U_{jk} + p Z_{ijk} \leq p - W_j \qquad \forall i,j,k \quad (12)$$

$$W_i \leq U_{ik} \leq p \qquad \forall i, k \quad (13)$$

$$\sum_{0 \leq i \leq n+1} \sum_{0 \leq j \leq n+1} (t_{ij} + s_i) Z_{ijk} \leq r \qquad \forall k \quad (14)$$

$$\sum_{0 \leq i \leq n+1} \sum_{0 \leq j \leq n} Z_{ijk-1} \geq \sum_{0 \leq i \leq n+1} \sum_{0 \leq j \leq n} Z_{ijk} \qquad k \geq 2 \quad (15)$$

$$Z_{iik} = 0 \qquad \forall i, k \quad (16)$$

$$X_{ij} \in \{0, 1\} \qquad \forall i, j \quad (17)$$

$$Y_i \in \{0, 1\} \qquad \forall i \quad (18)$$

$$Z_{ijk} \in \{0, 1\} \qquad \forall i, j, k \quad (19)$$

$$U_{ik} \geq 0 \qquad \forall i, k \quad (20)$$

$$W_i \geq 0 \qquad \forall i \quad (21)$$

The objective function (1) minimizes the overall cost, which has two components: direct mobile clinic operation costs and other outreach trip related costs. Constraints (2) ensure that a mobile clinic can only serve locations within the MCD of the clinic. Constraints (3) ensure that a location is only assigned to an established mobile clinic. Constraints (4) ensure that each location is assigned to a mobile clinic. Constraints (5) compute the total vaccine volume handled at a mobile clinic based on the population that the clinic covers. These four sets of constraints define a typical facility location problem.

The next set of constraints relate to the vehicle routing problem. Note that node 0 denotes the origin and node ($n$+1) is the final node at the end of a trip; both represent the depot. Constraints (6) and (7) imply that each outreach trip departs from the depot ($i$=0) exactly once, while Constraints (8) and (9) imply that each outreach trip enters back into the depot ($i$=$n$+1) exactly once. Constraints (10) ensure that the flow that enters and departs any location $i$ is balanced in each outreach trip $k$. Constraints (6) – (10) thus ensure that every outreach trip is indeed a (0)-($n$+1) path.

Constraints (11) state that if there is a mobile clinic at this location (i.e., $Y_i = 1$), the vehicle enters and departs the location exactly once across all trips during a planning period. Constraints (12) are the vehicle-specific version of the MTZ subtour elimination constraints introduced by Miller, Tucker, and Zemlin [29]. Note that for a particular vehicle trip $k$ in which $j$ follows $i$, $Z_{ijk} = 1$ implies $U_{jk} \geq U_{ik} + W_j > U_{ik}$. Suppose there exists a subtour $(i, j, \ldots i)$, with $i \neq 0, n$. Then, $U_{jk} > U_{ik} > U_{jk}$ will lead to a contradiction. The first inequality in Constraints (13) ensures that a vehicle carries enough vaccine for each mobile clinic, while the second inequality ensures that the vehicle capacity is not exceeded. Note that if location $i$ is not a part of trip $k$ the values of $U_{ik}$ are irrelevant to the problem as long as they satisfy (13).

Constraints (14) state that the sum of the travel times and service times in a trip cannot be larger than the MTD. Constraints (15) are added to avoid degeneracy by ensuring that trip $k$ is never utilized if trip $k$-1 is not utilized; with these constraints we reduce the search space by making sure that vehicle trips are chosen in a sequence of 1, 2, 3... In addition, it ensures that outreach trips with more stops will have a lower index value. Constraints (16) – (21) are self-explanatory.

## 4. A Multi-period Modeling Approach

In Sections 2 and 3 we introduced a model that assumed all parameters are constant and deterministic, and the model is solved once. However, conditions in many targeted locations are not always stable and predictable. It can often be difficult to obtain accurate estimates of all problem parameters ahead of time, and these might change as we get closer to the implementation of the outreach trips. For example, because demand is a function of population and birth rate, it can be more accurate to think of it as being stochastic, as both the population and the birth rate within a location could vary from year to year or even within a year. Similarly, in assumption (7) we estimate the travel time from $i$ to $j$ as a constant based on the distance and the average vehicle speed. However, traffic and road conditions in the targeted zones are often unstable, so that this assumption might also need to be reconsidered. With an extreme event such as a flood or a landslide, a road might even be blocked. Conversely, improvements to infrastructure might reduce travel times. Therefore, it would in general be suboptimal to determine a fully fixed strategy ahead of time and simply repeat it in every planning period.

On the other hand, it can also be problematic if we update all parameters and obtain completely revised plans for each planning period. Recall that our problem is to minimize costs while providing the *opportunity* for 100% coverage; however, for a variety of reasons, in practice not every patient will show up at a clinic. A major goal of the WHO is to make access to vaccinations as easy as possible so as to minimize the number of these so-called lost opportunities. From this viewpoint, it is desirable to have a stable set of locations for mobile clinics and for the populations assigned to each to be aware of where and when clinics will be conducted on a regular basis (e.g., the second Tuesday of every month; the first Monday in January, April, July and October; March 15 and September 15). We draw a compromise here by fixing locations but allowing for flexibility in timings during different planning periods. It is undesirable to move locations where mobile clinics are conducted because it is disruptive and confusing for the populace to be directed to a different location each time for vaccination services. In contrast, it is relatively easy to inform people of a change in the timing of a mobile clinic (that might arise because of a change in how we do the vehicle routing), especially if it is only for some clinics and the new times are not too different from those in the previous planning period.

Suppose we consider our problem using a multi-period stochastic modeling approach. The two main uncertainties we consider during each planning period are (a) with respect to the population (and hence the volume of vaccines required) at each location, and (b) with respect to the travel times between locations $i$ and $j$. Suppose that the volume of vaccines demanded at location $i$ within each planning period is stochastic and represented by the random variable $\tilde{b}_i$, but we can constrain it to lie within some range $(\underline{b_i}, \overline{b_i})$. Similarly, the travel time between $i$ and $j$ is assumed to be stochastic and given by $\tilde{t}_{ij} \in (\underline{t_{ij}}, \overline{t_{ij}})$. This range might in general, be as wide as desired during the initial planning period in order to account for inherent uncertainties and the upper bounds would reflect the worst-case scenarios. We can then utilize MIP-1 but incorporate this information to now minimize either the *expected* cost or the *maximum* possible cost. We choose the latter option as it is more conservative and aligns better with LMIC planners who have to follow the WHO guidelines of reaching every child. As we will see, this also has the advantage of not requiring a characterization of the probability distributions associated with the stochastic variables, which would be virtually impossible to obtain. The solution to this problem yields (1) the optimal locations for mobile clinics along with the assignment of other locations to these mobile clinics, and (2) the associated routes for outreach trips for use within the initial planning period. The first set of decisions will apply to all subsequent planning periods, but the operational decisions on outreach trips and which location forms part of which trip will be updated during each subsequent planning period.

At the end of each planning period we review our estimates of the demand and travel time parameters and update these based on the most current information. For example, estimates of the population in some locations might have changed because of seasonal migrations or because of updated information from public health or other sources. Similarly, we might perhaps know that because of some natural catastrophe certain roads will be unavailable over the next planning period, or that driving times along certain routes will be longer or shorter because of changes in the season or changes in road conditions.

In the next planning period problem, we assume that the locations of mobile clinics and the locations assigned to be covered by each clinic are fixed based on the solution to the problem at the initial planning period. However, we use updated parameters $(\underline{b_i'}, \overline{b_i'})$ and $(\underline{t_{ij}'}, \overline{t_{ij}'})$ to obtain revised routes for outreach trips in this planning period. This process is repeated for each new planning period.

We illustrate this via a simple example shown in Figure 1. Suppose that we are developing the outreach strategy for an area containing a fixed clinic (which serves as the depot) and 15 locations that it must cover via outreach within each planning period of three months. We use our initial estimates of the demand and travel time to obtain the locations for mobile clinics, along with the assignments of other locations to each clinic. We also obtain a set of outreach trips with routes as shown in Figure 1, where arrows represent vehicle routes and dotted lines represent assignment of locations without mobile clinics to a particular mobile clinic. Here we will conduct 8 mobile clinics at locations 2, 5, 6, 9, 10, 12, 14, 15, spread across three separate outreach trips: trip 1 visits and holds mobile clinics at locations 2, 5 and 6 before returning to the depot; trip 2 does the same with locations 10 and 9, and trip 3 with locations 12, 14 and 15. While each mobile clinic serves the population at its location, the clinic at location 2 also serves locations 3 and 4, which are within the MCD of location 2. Similarly location 6 also serves 7; 9 also serves 8; 12 also serves 11 and 13; and people at location 1 are served by the depot (location 0).

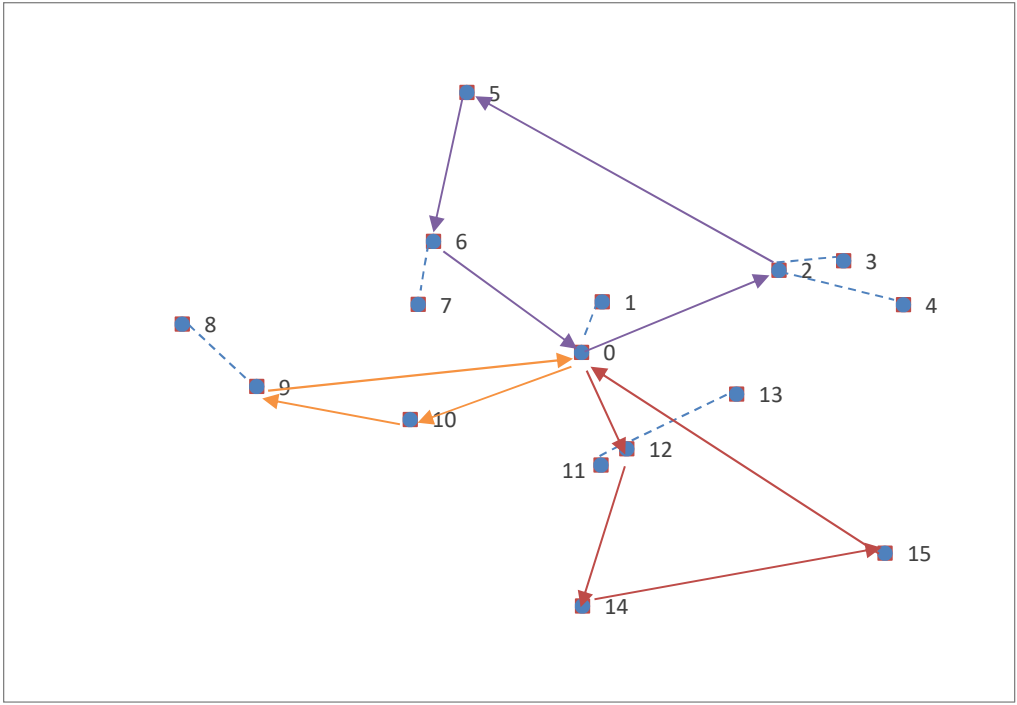

*Figure 1: Initial Solution*

This plan is implemented in the first planning period (quarter, in this case). At the end of the planning period we get updated information and learn that the travel time along each edge during the next planning period will be a lot shorter because we have a new vehicle now, but that the roads connecting locations 2 and 5, as well as 0 and 6 will be closed because of major repairs. Without changing the locations of our mobile clinics and the locations assigned to be covered by each clinic, we would like to obtain a possibly better set of outreach trips to cover these same locations during the next planning period based on this updated information. This results in the strategy displayed in Figure 2. We still have our mobile clinics at the same eight locations with the same assignments, but we now have only two outreach trips (0-12-14-15-2-0 and 0-5-6-9-10-0).

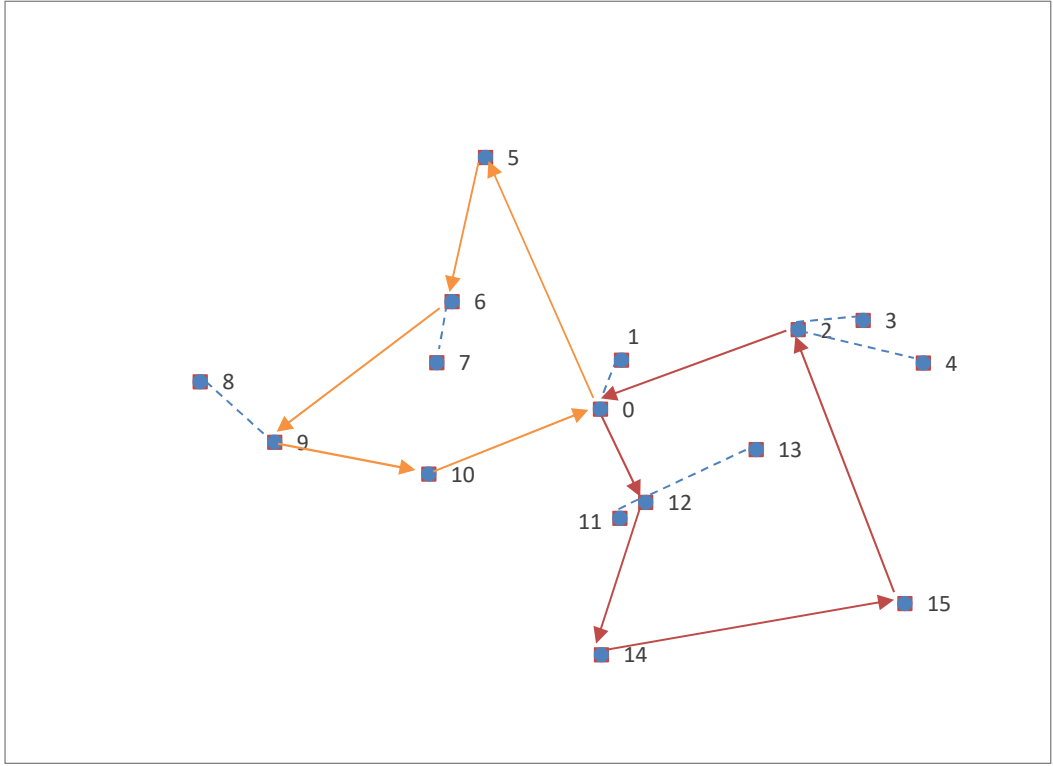

*Figure 2: Updated Solution*

This process can then be repeated for subsequent planning periods, and as we obtain new information we can obtain updated solutions for the outreach trips each time. In summary, we solve the following stochastic MIP problem:

**S-MIP-1:**

$$Z_1 = Min_{X,Y,Z,U,V,W} Max \sum_{1 \leq i \leq n} f_i Y_i + \sum_{0 \leq i \leq n+1} \sum_{0 \leq i \leq n+1} \sum_{1 \leq k \leq K} c\tilde{t}_{ij} Z_{ijk} \qquad (22)$$

$subject\ to$

Constraints (2) – (4)

$$W_i = \sum_{0 \leq j \leq n+1} \tilde{b}_j X_{ij} \qquad i \leq n \quad (23)$$

Constraints (6) – (13)

$$\sum_{0\leq i\leq n+1}\sum_{0\leq j\leq n+1}(\tilde{t}_{ij}+s_i)Z_{ijk} \leq r \qquad \forall k \quad (24)$$

Constraints (15) – (21)

$$\tilde{t}_{ij} \in (\underline{t_{ij}}, \overline{t_{ij}}); \tilde{b}_i \in (\underline{b_i}, \overline{b_i})$$

Note that the constraints (23) and (24) replace (5) and (14) in MIP-1 by defining vaccine volumes and travel times to be stochastic, but bounded as above.

Let $X_{ij}^*$ and $Y_i^*$ be the optimal value of $X_{ij}$ and $Y_i$ in the solution to S-MIP-1. Then we define:

**S-MIP-2:**

$$Z_2 = \left(\sum_{1\leq i\leq n} f_i Y_i^*\right) + Min_{Z,U,V,W}\ Max \sum_{0\leq i\leq n+1}\sum_{0\leq j\leq n+1}\sum_{1\leq k\leq K} c\tilde{t}_{ij}Z_{ijk} \qquad (25)$$

subject to

Constraints (2) – (4), with $X_{ij} = X_{ij}^*$ and $Y_i = Y_i^*$

$$W_i = \sum_{0\leq j\leq n+1} \tilde{b}_j X_{ij}^* \qquad i \leq n \quad (26)$$

Constraints (6) – (13), with $Y_i = Y_i^*$ in (11)

$$\sum_{0\leq i\leq n+1}\sum_{0\leq j\leq n+1}(\tilde{t}_{ij}+s_i)Z_{ijk} \leq r \qquad \forall k \quad (27)$$

Constraints (15) – (17), (19) – (21)

$$\tilde{t}_{ij} \in (\underline{t'_{ij}}, \overline{t'_{ij}}); \tilde{b}_i \in (\underline{b'_i}, \overline{b'_i})$$

This problem is identical to S-MIP-1, but with the locations of the mobile clinics and the locations assigned to be covered by each clinic now fixed at those obtained from the solution to S-MIP-1, so that we are only solving for an optimal set of trips.

At the beginning of the initial planning period, we solve S-MIP-1 with the initial estimated ranges $(\underline{b_i}, \overline{b_i})$ and $(\underline{t_{ij}}, \overline{t_{ij}})$ that are as wide as desired to account for inherent uncertainties and with upper bounds that reflect the worst-case scenarios. The solution to this $(X_{ij}^*, Y_i^*, Z_{ijk}^*)$ yields the optimal locations for mobile clinics $(Y_i^*)$, the assignment of other locations to these $(X_{ij}^*)$, and the associated routes for outreach trips $(Z_{ijk}^*)$ that are to be used during the initial planning period. At the end of the period, we get updated information $(\underline{t_{ij}'}, \overline{t_{ij}'})$ and $(\underline{b_i'}, \overline{b_i'})$. Without changing the locations of our mobile clinics that are defined by $Y_i^*$ (obtained from S-MIP-1) and the assignment of locations to mobile clinics defined by $X_{ij}^*$ (also obtained from S-MIP-1), we find a possibly better set of outreach trips to cover these same locations during the next planning period. We repeat the process of solving S-MIP-2 for each subsequent planning period with updated demand and travel time parameters, while using the same locations for mobile clinics as defined by $Y_i^*$ and the assignment as defined by $X_{ij}^*$. Of course, if at some point it becomes appropriate or necessary to change locations for mobile clinics, we can go back and solve S-MIP-1 again to restart the process.

**Proposition 1**: Assuming feasibility, S-MIP-1 and S-MIP-2 are equivalent respectively, to (a) solving S-MIP-1 with $\tilde{t}_{ij} = \overline{t_{ij}}$ in (22), (24) and $\tilde{b}_j = \overline{b_j}$ in (23); and (b) solving S-MIP-2 with $\tilde{t}_{ij} = \overline{t_{ij}'}$ in (25), (27) and $\tilde{b}_j = \overline{b_j'}$ in (26).

<u>Proof</u>: First, note that from (23) or (26), as the value of $\tilde{b}_j$ increases, so does the value of $W_i$. This in turn reduces the size of the feasible regions for S-MIP-1 and S-MIP-2 by tightening the constraints defined by (12) and (13). Similarly, an increase in $\tilde{t}_{ij}$ tightens the constraints defined by (24) or (27) while also increasing the cost coefficient for $Z_{ijk}$ in the objective. So with these changes, assuming feasibility, the objective function can only increase from its current value (or at best, stay the same). Its maximum value is thus obtained when each $\tilde{b}_j$ and $\tilde{t}_{ij}$ is at its largest possible value.

The above result is intuitive: when the population (demand) increases, it is possible that limitations arising from the vehicle capacity and larger service times might increase the number of trips required to cover all locations. When travel times along a link $i$-$j$ increase, the total travel costs rise; it is also possible that the length of a trip might exceed the trip MTD ($r$), again causing an increase in the number of trips. Proposition 1 states that if we are conservative and plan for the worst-case scenario with respect to subsequent planning periods, then this corresponds to when travel times and populations are as large as they could get. We refer to the solutions for these worst-case scenarios as *robust* solutions.

## 5. Robustness and the Value of Information

In Section 4 we introduced a multi-period procedure to address the unstable outreach environment that is typical in practice, and to obtain robust solutions. In this section we compare, discuss and interpret the costs associated with the robust solutions for the initial planning period and subsequent planning periods.

We can interpret $Z_1$, the optimal value of S-MIP-1 as the optimal cost associated with the conservative strategy at the beginning of the initial planning period that addresses the worst-case scenario. The optimal value of S-MIP-2 given by $Z_2$ is also for a conservative strategy, but one that has an updated worst-case scenario and has clinic locations and the assignment of locations to each fixed based upon the optimal solution to S-MIP-1. Any difference between $Z_1$ and $Z_2$ is a result of possibly updated outreach trips with better vehicle routes. Note that $Z_2$ could in general be larger or smaller than $Z_1$. However, if the updated upper bounds are the same or smaller than before, then as the following corollary states, $Z_2$ will always be smaller.

**Corollary 1**: If $\overline{b'_i} \leq \overline{b_i}$ and $\overline{t'_{ij}} \leq \overline{t_{ij}}$, then $Z_2 \leq Z_1$.

<u>Proof</u>: In proving Proposition 1 we saw that as the values of $\tilde{b}_i$ and $\tilde{t}_{ij}$ increase, the feasible regions for both problems shrink, and when they decrease the region expands. Therefore, $Z_1$ and $Z_2$ are monotone non-decreasing in both $\tilde{b}_{ij}$ and $\tilde{t}_{ij}$. Further, S-MIP-2 has the same locations as the optimal locations in S-MIP-1 (at $i$ corresponding to $Y_i^* = 1$), and if $\overline{b'_i} \leq \overline{b_i}$ and $\overline{t'_{ij}} \leq \overline{t_{ij}}$ it has an expanded feasible region for choosing the delivery routes; so $Z_2 \leq Z_1$.

**Definition 1:** The percentage improvement in the robust cost that arises from tighter upper bounds is defined as $\Delta Z = 100 * (Z_1 - Z_2)/Z_1$.

Note that the locations for mobile clinics used in S-MIP-2 were obtained by solving S-MIP-1, and in general, these need not be optimal with the updated problem parameter estimates. If we had the ability to relocate mobile clinics in each planning period, we could design a network and an associated outreach strategy with a possibly lower cost than $Z_2$ for the new planning period. Note that this would be the best strategy for the new planning period if we were free to completely re-optimize the entire outreach system for each planning period. To this end, we define the following Optimal Stochastic MIP model (OS-MIP-2) whose value $Z_0$ is the smallest worst-case cost incurred with updated information, but when we have flexibility with respect to the choice of locations for mobile clinics and the assignment of locations to each (as opposed to S-MIP-2, which has no flexibility to change these locations).

**OS-MIP-2:**

$$Z_0 = Min_{X,Y,Z,U,V,W} \, Max \sum_{1 \leq i \leq n} f_i Y_i + \sum_{0 \leq i \leq n+1} \sum_{0 \leq j \leq n+1} \sum_{1 \leq k \leq K} c\tilde{t}_{ij} Z_{ijk} \quad (28)$$

subject to

Constraints (2) – (4)

$$W_i = \sum_{0 \leq j \leq n+1} \tilde{b}_j X_{ij} \qquad \text{for } \forall i \leq n \quad (29)$$

Constraints (6) – (13)

$$\sum_{0 \leq j \leq n+1} \sum_{0 \leq j \leq n+1} (\tilde{t}_{ij} + s_i) Z_{ijk} \leq r \qquad \text{for } \forall k \quad (30)$$

Constraints (15) – (21)

$$\tilde{t}_{ij} \in (\underline{t'_{ij}}, \overline{t'_{ij}}); \tilde{b}_i \in (\underline{b'_i}, \overline{b'_i})$$

Note that this problem is identical to S-MIP-1 but with bounds on the times and the demands that correspond to the ones used in S-MIP-2 for the new planning period. Furthermore, it differs from S-MIP-2 in that although the bounds are the same, $X_{ij}$ and $Y_i$ are decision variables here, while in S-MIP-2 these were fixed at $X_{ij}^*$ and $Y_i^*$ obtained from solving S-MIP-1.

Along similar lines as Proposition 1 we have:

**Proposition 2**: Program OS-MIP-2 is equivalent to solving it with $\tilde{t}_{ij} = \overline{t'_{ij}}$ in (28), (30) and $\tilde{b}_j = \overline{b'_j}$ in (29).

The following proposition relates OS-MIP-2 to S-MIP-2:

**Proposition 3**: $Z_0 \leq Z_2$.

Proof: It is clear that OS-MIP-2 is a relaxation of S-MIP-2, with the option of picking locations other than those given by $Y_i = Y_i^*$ and assignments other than $X_{ij} = X_{ij}^*$. Therefore, $Z_0 \leq Z_2$.

Note that the optimal value of OS-MIP-2 ($Z_0$) is from yet another conservative objective, and corresponds to the theoretical best robust solution to the outreach problem for the new planning period. Any difference between $Z_2$ and $Z_0$ is due to the fact that in OS-MIP-2 we have the freedom to update locations for mobile clinics and assignments. We may also interpret this difference as the value of having better information on the parameter bounds at the beginning of the initial planning period, as opposed to having to wait for it until the beginning of the new planning period (because we would then have obtained this solution when solving S-MIP-1 for the first period).

**Definition 2:** The value of information is defined as $V = 100 * (Z_2 - Z_0)/Z_2$.

Thus $V$ is the percentage savings possible (in the worst-case scenario), from obtaining information in the form of correct bounds on $\tilde{t}_{ij}$ and $\tilde{b}_i$ at the beginning of the first period.

## 6. Numerical experiments

We tested the procedure described in the previous sections on data that we adapted from four countries in sub-Saharan Africa. The goals of our numerical experiments were to (a) see how operational plans might change with updated information, and (b) to gain managerial insight on the relative importance of demand information and travel time information, and how these interact with demographic characteristics. Due to issues with data confidentiality we label these countries A through D. Country B is relatively small but with a high population density. The other three are larger in area and have some pockets of dense population, with others (such as areas in the Saharan desert) that are much more sparsely populated. To explain our numerical experiments and demonstrate some of the insights to be gained, we will first describe in detail an illustrative example with plans derived for the initial planning period and the next (new) planning period, using data derived from Country D. Following this, we analyze and summarize results from a larger set of instances. All of our computations were done using a computer with an Intel Core i5

CPU and a 2.80 GHz processor with 8.0 GB memory. For solving the integer programs, we used a standard commercial solver (Gurobi v8.1.0rc1).

Our illustrative example has 9 locations on a 20 km by 20 km graph, with the depot being in the middle of the graph, and with each location having an average of 100 newborns in a year. For implementation in the initial planning period a robust solution (with value $Z_1$) is obtained for problem S-MIP-1 using initial estimates of upper bounds on demand and travel times. The locations of mobile clinics and the locations assigned to each clinic in this solution are then fixed. Next, using the most current information on demands and travel times, vehicle routes are updated for the next planning period by obtaining a robust solution (with value $Z_2$) to problem S-MIP-2. We also solve OS-MIP-2 to obtain the theoretical best robust solution to the problem for the new period (with value $Z_0$). We study the impact of upper bound changes (i) only in $b$ (demand), (ii) only in $t$ (travel times), and (iii) in both $b$ & $t$ under the assumption that the bounds on the updated estimates are tighter than the initial ones.

We first generated a base case for the new planning period with associated values for $\overline{b'_j}$ and $\overline{t'_{ij}}$, and obtained $Z_0$ by solving problem OS-MIP-2. This solution, with a value of $Z_0$= \$619.17 for our example, represents the best robust solution obtainable for the new planning period and is illustrated in Figure 3: it has four clinics (at locations 1, 2, 3, and 5) with two trips (0-1-2-0) and (0-3-5-0). The mobile clinics at locations 2, 3 and 5 also cover the populations at 7, (4,6) and (8,9) respectively. The cost of \$619.17 is comprised of \$290.50 in direct costs to run mobile clinics and \$328.67 in other costs associated with outreach trips.

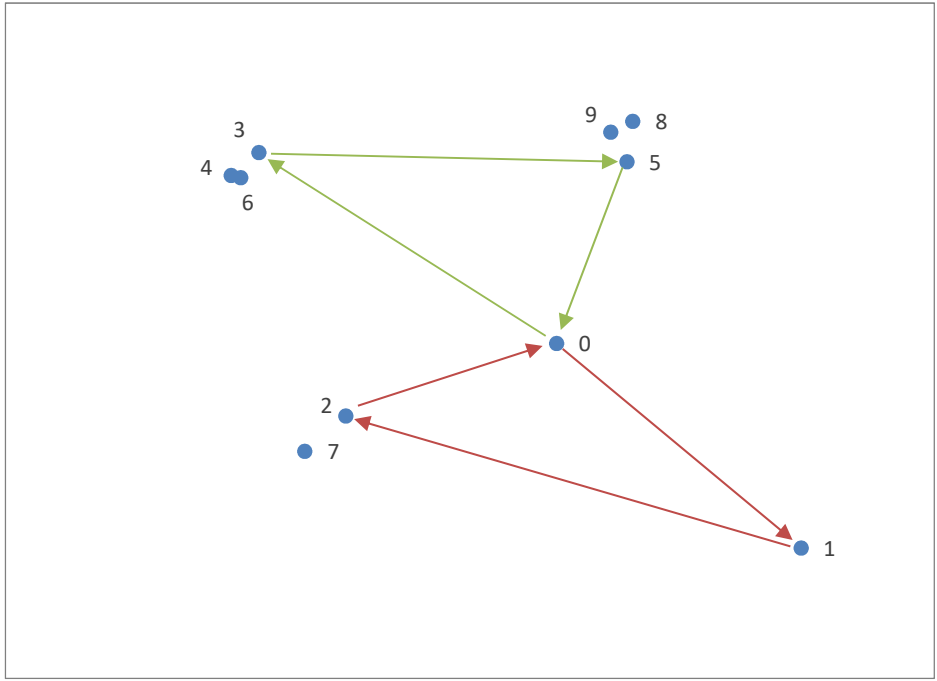

*Figure 3: Optimal Robust Solution for the New Planning Period Problem*

Next, for each of the three types of parameter changes ($b$, $t$, and $b$ & $t$) we studied (a) small, (b) moderate and (c) large reductions in the initial estimates of the upper bounds before the first planning period (from $\overline{b_j}$ and $\overline{t_{ij}}$ to $\overline{b'_j}$ and $\overline{t'_{ij}}$, respectively). Specifically, for these three cases we assumed that $\overline{b_j}$ and $\overline{t_{ij}}$ were on average 20%, 80% or 150% larger than $\overline{b'_j}$ and $\overline{t'_{ij}}$. In all cases, we first solve problem S-MIP-1 with the appropriate values of $\overline{b_j}$ and/or $\overline{t_{ij}}$ to obtain the robust solution for the initial planning period, along with its value $Z_1$. We then fix clinic locations and their allocations to solve problem S-MIP-2 using $\overline{b'_j}$ and $\overline{t'_{ij}}$ for the bounds, and obtain the robust solution for the new planning period, along with its value $Z_2$. The results are listed in Table 1, and we discuss some of the insights that these offer.

*Table 1: Example in Country D*

|  | b | | | t | | | b & t | | |
|---|---|---|---|---|---|---|---|---|---|
| Case | Small | Moderate | Large | Small | Moderate | Large | Small | Moderate | Large |
| $Z_1$ | 619.17 | 619.17 | 1003.68 | 651.29 | 714.19 | 778.49 | 693.39 | 831.13 | 1682.26 |
| $Z_2$ | 619.17 | 619.17 | 817.49 | 630.32 | 630.32 | 642.03 | 619.35 | 627.30 | 923.71 |
| $\Delta Z$ | 0.00% | 0.00% | 18.48% | 3.22% | 11.74% | 17.53% | 10.68% | 24.52% | 45.09% |
| V | 0.00% | 0.00% | 24.26% | 1.77% | 1.77% | 3.56% | 0.03% | 1.29% | 32.97% |

Note: All costs are in $. Theoretical best robust optimum for the new planning period = $Z_0$ = **$619.17**

First, it may be seen that with the tighter bounds, the robust optimum for S-MIP-2 (=$Z_2$) shows improvement over that for S-MIP-1 (=$Z_1$) in seven of the nine cases, with the percentage improvement ($\Delta Z$) being much more significant when the upper bounds get tighter (i.e., with larger reductions). While these results are intuitive, it is interesting that the improvements are more pronounced with tighter time estimates as compared to tighter demand estimates (and simultaneous reduction of uncertainty in both parameters further magnifies the savings).

Next, we look at the issue of what we could have achieved in the new planning period if we had been free to re-optimize locations and assignments. That is, we compare the robust optimum $Z_2$ from S-MIP-2 to its theoretical lowest value of $Z_0$ =$619.17. In particular, we compute the theoretical maximum percentage improvement possible in $Z_2$, i.e., the value of information ($V$) as given by Definition 2. It may be observed that for this example, the updated robust solution to the new planning period problem is actually very close to the theoretical best value in seven of the nine cases. The only instances where the value of information is high is when there are large reductions in the estimated upper bounds for demand, especially when there are

simultaneous reductions in the upper bound on travel times. We visually illustrate some of the results in Figures 4 to 7.

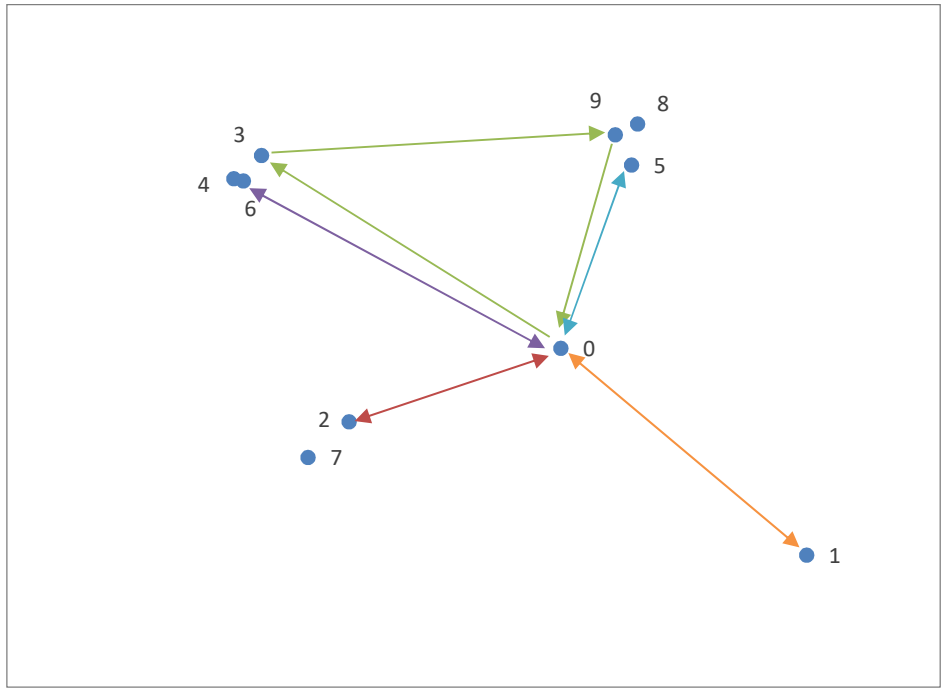

*Figure 4: Solution for the Initial Planning Period*

First, consider demand estimates. If the revised estimates in the upper bounds are only slightly or moderately tighter (columns 1 and 2 in Table 1), we find the theoretical best solutions at the beginning and this does not change with revised estimates; thus there is no value to these revised estimates. In contrast, if there is a large reduction in the estimate from the initial planning period to the new planning period (column 3 in Table 1) the situation is different. Figure 4 illustrates the solution to problem S-MIP-1 for the first planning period, with six clinics scheduled via five outreach trips covering 1, 2, (3, 9), 5, 6 respectively. Note that clinics at 2, 9 and 6 also cover the populations at locations 7, 8 and 4, respectively in this solution. As displayed in Table 1, this solution has a total cost of $1003.68.

Now, consider the new planning period where we are constrained to maintain outreach clinics at these same six locations, but use the updated information on the worst-case demand to solve problem S-MIP-2. This yields the updated solution shown in Figure 5 with two outreach trips (0, 1, 5, 9, 0) and (0, 3, 6, 2, 0) to cover the six clinics, and a total cost of $817.49. Locations 7, 8 and 4 are covered by the clinics at 2, 9 and 6, respectively. Note that the very loose initial upper bound for demand caused the robust solution to the initial planning period to have more trips because a potentially much larger demand could cause vehicle capacity constraints to be violated if the new planning period solution is adopted. In both solutions we have facility costs of $435.75 for the six open mobile clinics, but trip costs of $567.93 in the initial planning period as opposed to $381.74 in the new one.

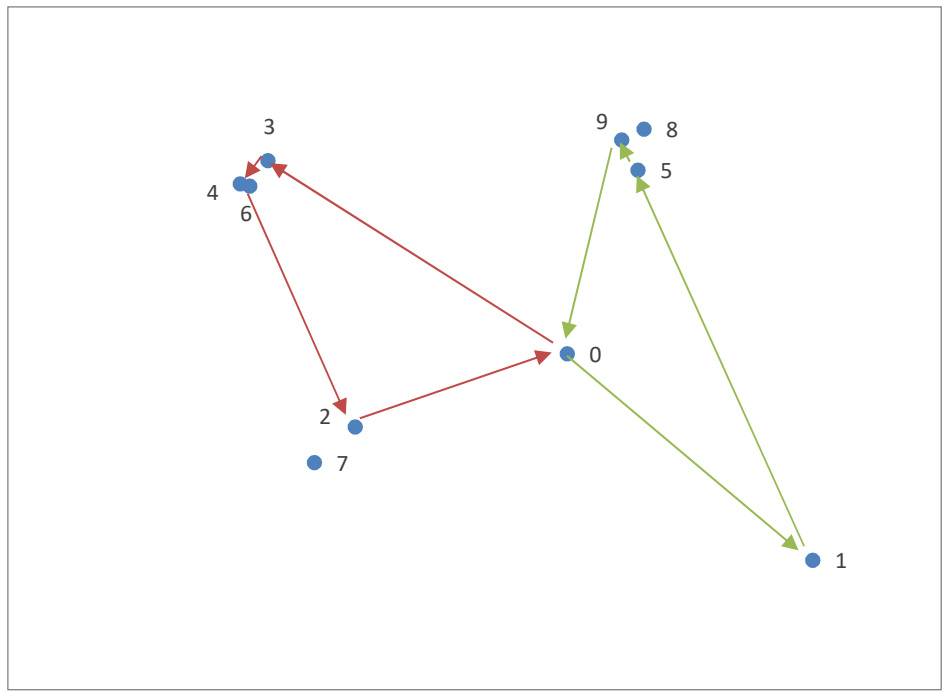

*Figure 5: Updated Solution in the New Period with Large Reductions in Demand Estimates*

Finally, note that if we had had the updated information prior to the initial planning period, we would have obtained the best overall robust solution (shown in Figure 3), which is 24.26% better than the one from S-MIP-2; this is the value of information in this instance.

Next, we look at travel time; Figure 6 is for the case where upper bounds on travel times are tightened slightly or moderately, i.e., when the initial upper bound estimates were either 20% or 80% larger on average than their updated values in the new planning period (columns 4 and 5 in Table 1). The robust solutions for both planning periods are identical in both cases, and the reductions from $Z_1$ to $Z_2$ ($651.29 to $630.32, and $714.19 to $630.32, respectively) are only because $\overline{t'_{ij}} < \overline{t_{ij}}$. Also, the only difference between Figure 6 and the theoretical best scheme for the new planning period shown in Figure 3 is that a clinic is located at 8 instead of 5. This is because $\overline{t_{05}}$ happens to be larger than $\overline{t_{08}}$, and thus in the initial planning period, location 8 is preferable to location 5 in the solution to S-MIP-1. When in the new planning period, the clinic is fixed at location 8 and we use $\overline{t'_{05}}$ and $\overline{t'_{08}}$ in Problem S-MIP-2, the cost ($Z_2$=$630.32) is only $11.15 higher than it would have been ($Z_0$=$619.17) with the optimal location (i.e., $V$=1.77%).

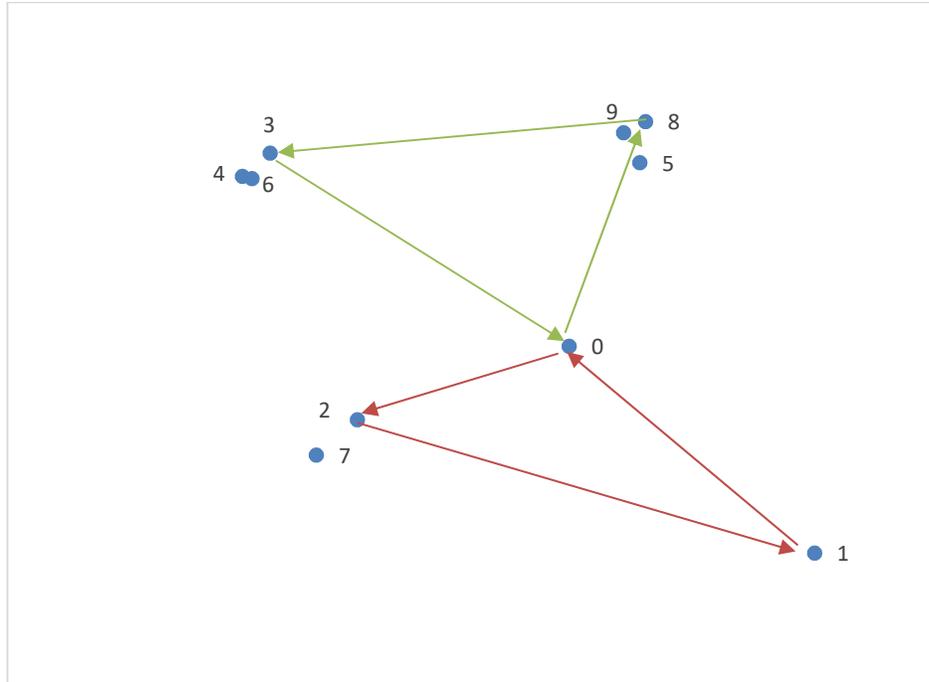

*Figure 6: Solutions in Both Planning Periods: Small or Moderate Reductions in Travel Time Estimates*

Lastly, Figure 7 depicts the case where there is a large reduction (column 6 in Table 1) in the initial estimates of travel time ($\overline{t_{ij}}$ exceeds $\overline{t'_{ij}}$ by 150% on average). Again, the solutions are identical in both planning periods, and in contrast with the case when reductions in the bounds are small or moderate, a clinic is now assigned to location 7 instead of location 2. Using these locations as opposed to the optimal ones in Figure 3 yield a cost of $Z_2$=$642.03 for the new planning period, that is $22.86 higher than the theoretical best value of $Z_0$ (with $V$ =3.56%).

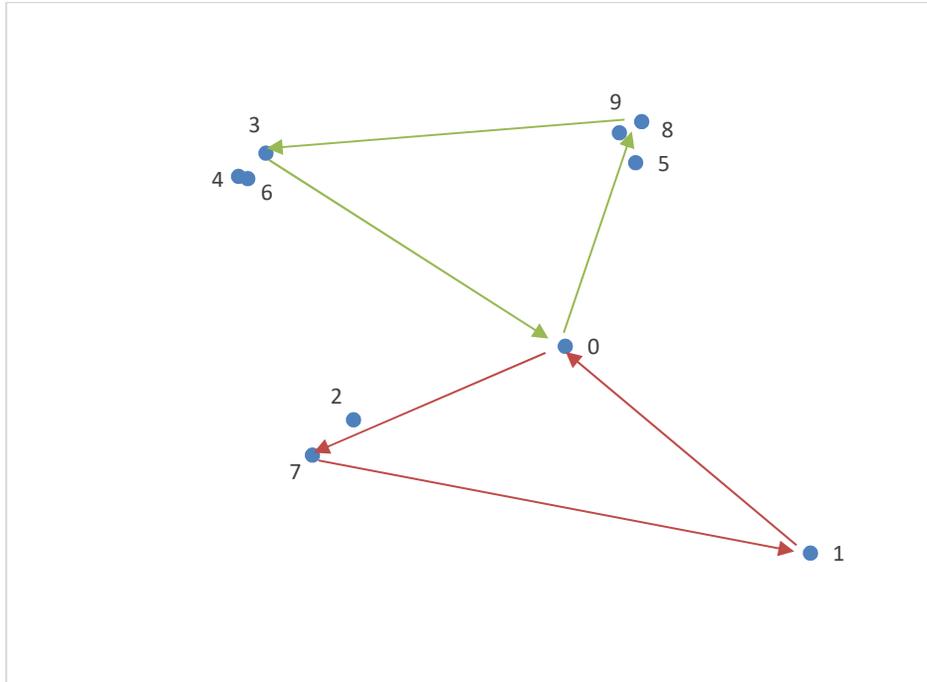

*Figure 7: Solutions in Both Periods: Large Reductions in Travel Time Estimates*

With simultaneous changes in demand and travel times parameter estimates (columns 7, 8 and 9 in Table 1) similar types of results are obtained; figures are omitted in the interest of brevity.

We now present results from a larger set of instances across all four countries using the same methodology. Without loss of generality, we only study and compare one new planning period with the initial planning period, since subsequent planning periods continue the identical process of using the same locations for mobile clinics and only updating bounds on demand and travel time.

The results for the value of information ($V$) are summarized in Tables 2, 3 and 4 based upon the demographic characteristics of the region. Table 2 shows results for examples from Country B and regions of Countries A, C, and D with high population densities (e.g., around their capital cities). Table 3 has instances from Countries A, C, and D where populations are moderately distributed, while Table 4 covers larger, often remote regions in the same countries, with relatively sparse populations. Note that the examples in Table 4 have fewer locations that are more sparsely distributed on a larger graph, while the examples in Table 2 have more locations on a relatively small graph; the examples in Table 3 are in between these two extreme cases. Also note that smaller values for $V$ in the tables indicate worst-case solutions that are relatively robust with our approach, while larger values indicate that having tighter, more accurate initial estimates of parameter bounds can result in more significant savings over the solutions from our approach.

We separately highlight instances with $V$ values over 20% and those with values under 5% to indicate "large" and "small" values, respectively for the value of information. The entries not highlighted may be thought of as being in between. The implications of the numbers in these tables and insights that can be drawn from them are discussed in detail in the next section.

*Table 2: Value of information for examples in smaller, densely populated regions*

|        | $b$   |          |        | $t$   |          |        | $b$ & $t$ |          |        |
|--------|-------|----------|--------|-------|----------|--------|-----------|----------|--------|
|        | Small | Moderate | Large  | Small | Moderate | Large  | Small     | Moderate | Large  |
| A1     | 9.72% | 17.71%   | 32.47% | 0.00% | 0.00%    | 0.00%  | 9.72%     | 17.71%   | 32.47% |
| B1     | 0.00% | 22.87%   | 50.99% | 0.38% | 1.96%    | 1.96%  | 0.38%     | 23.35%   | 50.99% |
| B2     | 0.00% | 15.47%   | 47.88% | 0.05% | 0.05%    | 0.05%  | 0.05%     | 15.47%   | 47.88% |
| B3     | 0.00% | 19.76%   | 49.24% | 0.00% | 1.09%    | 2.22%  | 0.00%     | 20.02%   | 49.24% |
| B4     | 31.22%| 31.70%   | 48.00% | 0.00% | 0.00%    | 31.87% | 31.22%    | 31.70%   | 58.58% |
| B5     | 0.00% | 17.45%   | 30.35% | 0.26% | 0.26%    | 0.26%  | 0.26%     | 17.51%   | 30.87% |
| B6     | 0.00% | 15.54%   | 26.85% | 0.00% | 0.00%    | 1.14%  | 0.00%     | 15.63%   | 36.13% |
| C1     | 24.57%| 38.75%   | 68.49% | 0.00% | 0.00%    | 0.00%  | 24.51%    | 38.82%   | 68.49% |
| C2     | 0.58% | 24.08%   | 39.33% | 0.58% | 1.75%    | 2.04%  | 0.58%     | 25.19%   | 50.24% |
| D1     | 20.05%| 41.38%   | 65.20% | 0.27% | 0.27%    | 2.62%  | 20.05%    | 42.57%   | 65.20% |
| Mean   | 8.61% | 24.47%   | 45.88% | 0.15% | 0.54%    | 4.22%  | 8.68%     | 24.80%   | 49.01% |
| Median | 0.29% | 21.32%   | 47.94% | 0.02% | 0.15%    | 1.55%  | 0.48%     | 21.68%   | 49.74% |

*Table 3: Value of information for examples in moderately populated regions*

|  | b | | | t | | | b & t | | |
|---|---|---|---|---|---|---|---|---|---|
|  | Small | Moderate | Large | Small | Moderate | Large | Small | Moderate | Large |
| A2 | 0.00% | 17.98% | 26.23% | 0.01% | 30.98% | 26.23% | 0.01% | 26.23% | 26.23% |
| A3 | 0.00% | 0.00% | 16.21% | 0.00% | 0.00% | 0.00% | 0.00% | 0.00% | 16.21% |
| A4 | 0.00% | 0.14% | 30.88% | 0.00% | 0.00% | 0.00% | 0.00% | 10.83% | 30.88% |
| C3 | 0.30% | 0.30% | 10.84% | 0.30% | 1.73% | 2.02% | 0.30% | 0.30% | 18.59% |
| C4 | 0.00% | 9.52% | 9.52% | 0.00% | 0.00% | 8.95% | 0.00% | 9.52% | 19.79% |
| D2 | 0.00% | 11.69% | 20.94% | 0.00% | 0.00% | 0.00% | 0.00% | 12.12% | 35.22% |
| D3 | 0.00% | 0.00% | 24.26% | 1.77% | 1.77% | 3.56% | 0.03% | 1.29% | 32.97% |
| D4 | 11.52% | 11.52% | 20.67% | 0.00% | 0.00% | 20.67% | 11.52% | 11.52% | 39.95% |
| D5 | 0.00% | 13.14% | 22.51% | 0.00% | 0.00% | 0.00% | 0.00% | 13.14% | 31.45% |
| D6 | 0.00% | 0.00% | 12.19% | 1.69% | 0.00% | 2.58% | 2.58% | 0.00% | 34.67% |
| Mean | 1.18% | 6.43% | 19.42% | 0.38% | 3.45% | 6.40% | 1.44% | 8.49% | 28.60% |
| Median | 0.00% | 4.91% | 20.80% | 0.00% | 0.00% | 2.30% | 0.00% | 10.17% | 31.16% |

*Table 4: Value of information for examples in larger, sparsely populated regions*

|  | b | | | t | | | b & t | | |
|---|---|---|---|---|---|---|---|---|---|
|  | Small | Moderate | Large | Small | Moderate | Large | Small | Moderate | Large |
| A5 | 0.00% | 0.00% | 0.00% | 2.42% | 2.02% | 2.02% | 2.42% | 2.02% | 17.99% |
| A6 | 0.00% | 0.12% | 0.12% | 0.12% | 8.42% | 8.42% | 0.12% | 8.42% | 8.42% |
| A7 | 0.00% | 6.98% | 14.29% | 0.00% | 14.29% | 14.29% | 0.00% | 14.29% | 14.29% |
| A8 | 0.00% | 0.00% | 16.00% | 0.00% | 0.00% | 0.00% | 0.00% | 0.00% | 16.00% |
| C5 | 6.81% | 6.81% | 6.81% | 0.28% | 0.00% | 0.00% | 6.56% | 6.56% | 17.78% |
| C6 | 0.00% | 0.00% | 7.97% | 0.00% | 0.00% | 0.00% | 0.00% | 7.97% | 15.40% |
| C7 | 0.00% | 5.51% | 5.51% | 0.00% | 0.00% | 5.51% | 0.00% | 5.51% | 11.57% |
| C8 | 0.00% | 0.00% | 0.00% | 0.00% | 0.00% | 0.00% | 0.00% | 0.00% | 0.00% |
| D7 | 0.00% | 0.00% | 4.91% | 2.05% | 4.91% | 4.91% | 2.05% | 4.91% | 11.25% |
| D8 | 0.00% | 0.00% | 0.00% | 0.66% | 0.96% | 0.96% | 0.00% | 0.96% | 0.96% |
| Mean | 0.68% | 1.94% | 5.56% | 0.55% | 3.06% | 3.61% | 1.11% | 5.06% | 11.37% |
| Median | 0.00% | 0.00% | 5.21% | 0.06% | 0.48% | 1.49% | 0.00% | 5.21% | 12.93% |

# 7. Discussion and Conclusions

First, consider changes in estimates of only $b$ or only $t$ (the first two columns in each of the tables). Although exceptions do exist, we can draw the general conclusion that our approach is quite robust (i.e., $V$ is small) in the following situations:

- Updates are only in travel time estimates; regardless of whether they are small, moderate or large the value of information is under 5% in 79, and over 20% in only 4 out of the 90 instances corresponding to this situation (the nine columns - across the three tables - under "$t$").

- Updates are only in the demand estimates and they are small: $V$ is under 5% in 24 and over 20% in only 3 out of the 30 instances for this case (the three columns under "$b$" and "Small").

- Updates are only in the demand estimates and they are moderate, but we are in larger areas with moderate to sparse population densities: $V$ is under 5% in 12 of 20 instances (the two columns under "$b$" and "Moderate" in Tables 3 and 4).

Conversely, the costs in the worst-case scenario can be significantly higher with our approach than they would be if we had perfect information in advance (i.e., $V$ is much larger) under the following scenarios:

- There are large updates in the bounds on demand: for this case, $V$ is over 20% in 16 out of the 30 instances and under 5% in only 5 out of the 30 instances (all for sparsely populated regions; the three columns under "$b$" and "Large").

- There are moderate updates in densely populated regions: $V$ is over 20% in 5 of the 10 instances for this case and never under 5% (column under "$b$" and "Moderate" in Table 2).

When we consider simultaneous changes in the estimated bounds for both travel time and demand estimates (the columns under "$b\ \&\ t$"), the results are closely correlated with what happens when there are changes in demand alone, leading one to conclude that demand updates constitute the primary factor and their effects overshadow updates in travel time estimates.

We also conducted a set of separate, nonparametric, Wilcoxon signed-rank tests for each of the three different types of regions ($d, m, s$) to see if there were differences in the magnitude of the mean effects of the different types of changes ($H_0: \mu_{\cdot b} = \mu_{\cdot t}$ vs. $H_1: \mu_{\cdot b} > \mu_{\cdot t}$ and $H_0: \mu_{\cdot b,t} = \mu_{\cdot b}$ vs. $H_1: \mu_{\cdot b,t} > \mu_{\cdot b}$ for each of $d, m, s$). Note that in each of the six comparison we have 30 paired instances across which we study difference in means. The null hypotheses were strongly rejected (P-values all under 0.001) for five out of these six tests; the only case where there was no significant difference in the mean value of $V$ was when comparing individual changes in estimates of $b$ and $t$ in sparse regions ($\mu_{sb}$ and $\mu_{st}$), which had a P-value of 0.66. In other words, (i) in dense and moderately populated regions, the value of information with updates only in demand is significantly higher (i.e., our approach is not as robust) as compared with updates only in travel times, and (ii) in all types of regions, the value of information is significantly higher with updates in both demand and travel times as compared with updates in demand alone.

Finally, we look at the effect of general demographic characteristics. When we have large, sparsely populated regions (Table 4) our approach is quite robust: $V$ is under 5% in 62 out of 90 instances, and always under 20%. When the population density is moderate (Table 3), our approach is still reasonably robust unless there are large changes in the estimated demand (as observed previously), in which case the value of $V$ starts to increase. Finally, in smaller, densely populated areas (Table 2 the results can be much more sensitive to changes in demand and there is significant value to obtaining more precise estimates of demand. Given that there are 90 instances for each type of region, we conducted two simple one-sided Z-tests for equality of means ($H_0: \mu_d = \mu_m$ vs. $H_1: \mu_d > \mu_m$ and $H_0: \mu_m = \mu_s$ vs. $H_1: \mu_m > \mu_s$). The null hypothesis is strongly rejected in favor of the alternative in both tests (P-values in both cases are under 0.0002), confirming that small, dense regions tend to have larger value of information than larger, moderately populated regions, and in turn, the latter yield larger value for $V$ than large, sparsely populated regions.

Based on our computational study, we can draw two main conclusions. First, larger sparsely populated regions tend to have lower value of information, while the opposite is true for smaller more densely populated regions. This means our approach is very robust in the former case, but less so in the latter case where there is value to obtaining tighter bounds on estimates. We speculate that this is because when there are fewer locations and they are relatively far apart and can serve relatively fewer neighboring locations, more locations are selected for mobile clinics, and capacity is less of an issue because fewer people are served by each outreach trip. Thus, even with perfect information, there is relatively little need to revise the initial plan even when demand and travel times estimates change. Conversely, in smaller, denser regions there are more dependencies between locations and more people are served in each trip. Thus changes in population estimates, and to a lesser extent, travel times, have a significant effect: often, the best strategy could be different from the plan that we obtain because capacities might be exceeded or alternative solutions to the set covering problem yield shorter vehicle trips. Thus the value of obtaining accurate information is much higher, and the solution with our approach might not be as robust.

Second, it is better to focus more attention on obtaining more accurate population (demand) estimates than travel time estimates. The latter have relatively low value of information and our approach is very robust even in smaller, densely populated regions with approximate estimates of these times. On the other hand, if demand estimates are too conservative (large) – as is commonly the case because of the goal of providing universal access - we could arrive at a strategy that results in locations that are not cost-effective after we get updated information; thus it is important to be able to get good estimates of demand in order for our approach to be robust.

In summary, this paper presents a systematic way to plan for economical outreach operations by formulating the problem as a mixed integer program. It also studies the issues related to the typical uncertainties associated with estimating demand for vaccines and planning individual outreach trips, and provides insights on where to focus attention if we are to follow a robust approach that plans for worst-case scenarios in order to comply with WHO-EPI guidelines to provide universal coverage.

## Declarations

**Funding**: This work was partially supported by the National Science Foundation via Award No. CMII-1536430.

**Conflicts of interest / Competing interests**: On behalf of all authors, the corresponding author states that there is no conflict of interest.

**Availability of data and material:** Not applicable.

**Code Availability**: Not applicable; standard commercial software (Gurobi)